\documentclass[conference]{IEEEtran}
\IEEEoverridecommandlockouts
\usepackage{cite}
\usepackage{amsmath,amssymb,amsfonts}
\usepackage{algorithmic}
\usepackage{graphicx}
\usepackage{textcomp}
\usepackage{xcolor}
\usepackage[acronym]{glossaries}
\usepackage{listings}
\usepackage{changepage} 
\usepackage[skins]{tcolorbox}
\usepackage{multirow}
\usepackage[normalem]{ulem}
\usepackage{enumerate}
\usepackage{xspace}

\def\BibTeX{{\rm B\kern-.05em{\sc i\kern-.025em b}\kern-.08em
    T\kern-.1667em\lower.7ex\hbox{E}\kern-.125emX}}
\begin{document}

\newacronym{APR}{APR}{Automated Program Repair}
\newacronym{FF}{FF}{File Frequency}
\newacronym{IAD}{IAD}{Inverse Average Distance to Crash Point}
\newacronym{IBF}{IBF}{Inverse Bucket Frequency}
\newacronym{JSF}{JSF}{Java Server Faces}
\newacronym{llm}{LLM}{Large Language Model}
\newacronym{RPD}{RPD}{Requests per day}
\newacronym{RPM}{RPM}{Requests per minute}
\newacronym{TPM}{TPM}{Tokens per minute (input)}
\newacronym{URI}{URI}{Uniform Resource Identifier}
\newacronym{MAP}{MAP}{Mean Average Precision}
\newacronym{ST}{ST}{stack trace}
\newacronym{SM}{SM}{stack message}
\newacronym{SCoSF}{SCoSF}{source codes of suspicious files}
\newacronym{RoSFM}{RoSFM}{rank of suspicious files and methods}
\newacronym{nSF}{nSF}{number of suspicious files}
\newacronym{nFF}{nFF}{number of files fixed}
\newacronym{nST}{nST}{number of stack traces}
\newacronym{nSM}{nSM}{number of stack messages}

\newcommand{\rqastbnumber}{RQ1}
\newcommand{\rqbstbnumber}{RQ2}
\newcommand{\rqastb}{\rqastbnumber\ -- What are the structural and explanatory characteristics of the responses generated by LLMs when suggesting fixes for software crashes?}
\newcommand{\rqbstb}{\rqbstbnumber\ -- What information provided for the clustering and ranking approach helps LLMs locate and fix the crash?}

\newcommand{\sds}{\textsc{SmallDS}\xspace}
\newcommand{\fds}{\textsc{FullDS}\xspace}
\newtcolorbox{rqanswer}[2]{%
  colback=gray!6,            
  colframe=black!80,         
  arc=3mm,                   
  boxrule=1pt,
  left=1pt,
  right=1pt,
  top=1pt,
  bottom=1pt,
  width=\columnwidth,
  boxed title style={
    colback=black!85,        
    colframe=black!85,
    arc=3mm,
    boxrule=0pt,
    left=1pt,
    right=1pt,
    top=1pt,
    bottom=1pt,
    width=\columnwidth
  },
  title={
    \color{white}\textbf{#1}
  }
}

\title{Integrating Crash Report Mining and LLMs for Bug Localization and Repair: An Industrial Report\\
}

\author{
    \IEEEauthorblockN{
        Marcos Medeiros\IEEEauthorrefmark{1}, Uirá Kulesza\IEEEauthorrefmark{1}, Christoph Treude\IEEEauthorrefmark{2}, Daniel Lucena\IEEEauthorrefmark{1}, Rafael Gomes\IEEEauthorrefmark{1}, \\ Roberta Coelho\IEEEauthorrefmark{1}, Eiji Adachi\IEEEauthorrefmark{1}, Rodrigo Bonifacio\IEEEauthorrefmark{3}
    }
    \IEEEauthorblockA{\small
        \IEEEauthorrefmark{1}\textit{Federal University of Rio Grande do Norte} (Natal, Brazil) \\ 
        Email: \{\texttt{marcosamm}, \texttt{rlucena.daniel}, \texttt{rafaggsique}\}@gmail.com, \{\texttt{uira}, \texttt{roberta}\}@dimap.ufrn.br, eijiadachi@imd.ufrn.br \\[0.5em]
        \parbox[t]{0.45\linewidth}{%
            \IEEEauthorrefmark{2}\textit{Singapore Management University} (Singapore, Singapore) \\
            Email: ctreude@smu.edu.sg
        }%
        \hfill
        \parbox[t]{0.45\linewidth}{%
            \IEEEauthorrefmark{3}\textit{Federal University of Pernambuco} (Recife, Brazil) \\
            Email: rbonifacio@cin.ufpe.br
        }
    }
}

\maketitle

\begin{abstract}
Analyzing crash-report bugs in large-scale industrial software systems requires substantial maintenance effort, particularly in production environments where developers must handle large volumes of crash reports and source code artifacts to localize and fix their root causes. While recent studies have shown that Large Language Models (LLMs) can assist with maintenance tasks, little is known about their effectiveness in supporting developers in analyzing crash-report bugs and repairing bugs associated with groups of crash reports in industrial settings. To address this gap, we investigate whether integrating crash report mining techniques---specifically stack trace clustering and suspicious file and method ranking---with LLMs can support crash localization and repair in production environments. We conduct a retrospective evaluation of five LLMs under four prompt configurations. After that, we chose the best model to run on 38 crash bugs collected from two large Java enterprise systems. We further analyze the structural characteristics and explanatory patterns of LLM-generated responses and assess localization and repair effectiveness through manual validation. Our results show that the best-performing configuration localizes up to 71\% and correctly repairs 52\% of crash bugs on the full dataset. These findings provide empirical evidence that combining crash report mining with LLM-based repair can effectively support debugging activities in industrial maintenance workflows.

\end{abstract}


\section{Introduction}

Crash reports and stack traces are widely used by developers to analyze system failures and support bug localization and resolution tasks~\cite{An2015, ken, Laura, bettenburg2008makes, schroter2010stack}. However, as crash reports accumulate over time, large-scale analysis becomes increasingly challenging~\cite{An2015, kinshumann2011debugging}. For example, Firefox receives millions of crash reports every month~\cite{Dhaliwal2011, ahmed2014impact}. In practice, developers must manually inspect stack traces and navigate large codebases to identify the root cause of a crash, which becomes increasingly costly as the volume of crash reports grows. To reduce this effort, prior research has investigated crash report aggregation techniques and shown that they can support bug localization and repair activities~\cite{podgurski2003automated, khomh2011entropy, kim2011crashes, dang2012rebucket, Wang2013, Wang2016, ball2003symptom, jones2002visualization, jones2005empirical, nessa2008software, schroter2010stack, wong2014boosting, gu2019does, Wu, wu2018changelocator}.

In parallel, the evolution and popularization of \glspl{llm} have enabled developers to leverage these systems for co-development and software quality assessment, particularly in program repair tasks~\cite{hou2024large, ferino2025junior}. Recent studies have explored the application of artificial intelligence to several software engineering tasks and report promising results~\cite{zhang2023critical, du2023resolving, du2026exploring, sobania2023analysis, fan2023automated, prenner2021automatic, prenner2022can, paul2023enhancing, jin2023inferfix}. 
Despite these advances, prior work~\cite{du2023resolving, zhang2023critical, du2026exploring} indicates that \glspl{llm} require sufficient contextual information and guidance to effectively locate and fix crash bugs. However, providing unrestricted access to entire codebases is often impractical in industrial environments due to cost, confidentiality, and LLM context window limitations. Thus, a key challenge is how to provide LLMs with structured and prioritized contextual information that narrows the search space while preserving relevance to the failure. 

Crash report mining techniques, such as stack trace clustering and suspicious file and method ranking, provide structured contextual information by identifying artifacts most likely associated with a crash. However, empirical studies conducted in industrial settings remain scarce~\cite{jarman2021legion, li2022empirical}. To the best of our knowledge, no previous work has systematically integrated crash mining with suspicious file and method ranking to provide structured contextual input to \glspl{llm} in a workflow validated on industrial production systems.

In this work, we present and empirically evaluate a pipeline that integrates crash report mining with \glspl{llm} to support crash bug localization and repair under realistic industrial constraints. The proposed workflow consists of three stages: crash report clustering, suspicious file and method ranking, and \gls{llm}-based patch suggestion. Our study focuses on large-scale web-based systems implemented using Java Enterprise technologies and adopts a retrospective evaluation to assess how effectively \glspl{llm} can locate and fix bugs from crash report data.
Our findings reveal that \glspl{llm} frequently suggest plausible fixes, even though structural heterogeneity makes automated patch integration infeasible. Surprisingly, the most explanation-prone model achieved the best localization and repair accuracy, while the most prompt-compliant model underperformed. 
The main contributions are:

\begin{itemize}
\item A novel approach that integrates crash report clustering and suspicious file and method ranking to provide structured contextual input for \glspl{llm} performing crash bug localization and repair.

\item An ablation-based evaluation on the impact of contextual information (stack traces, stack messages, and ranking information) on \gls{llm}-based crash localization and repair.

\item A validation of our approach that compares its ability to locate and fix crash root causes against developer corrective commits used as ground truth, showing that combining crash report mining with LLM-based repair effectively supports debugging in an industrial context.
\end{itemize}

The remainder of this paper is organized as follows. Section~\ref{sec:bug_location_and_fixing_approach_stb} presents the proposed bug localization and fixing approach. Section~\ref{sec:empirical_study_stb} describes the target systems, LLM models, prompt combinations, evaluation metrics, and study procedures. Section~\ref{sec:results_and_discussion_stb} presents and discusses the results, followed by threats to validity in Section~\ref{sec:threats_to_validity_stb}, related work in Section~\ref{sec:related_work_stb}, and conclusions in Section~\ref{sec:conclusion_stb}.

\section{Bug Localization and Fixing Approach} \label{sec:bug_location_and_fixing_approach_stb}

Our approach integrates existing crash report mining techniques with LLM-based repair suggestions into a unified debugging pipeline. Specifically, we adopt previously proposed methods for crash report grouping and suspicious artifact ranking, and use their outputs to construct structured contextual prompts for LLM-based bug localization and repair. The proposed approach consists of three main steps: (i) grouping crash reports that generate similar stack traces; (ii) ranking files suspected of causing the crashes; and (iii) using \glspl{llm} to localize bugs and suggest fixes for the code. The procedures for Steps (i)--(ii) were adopted from previous literature~\cite{medeiros2020, medeiros2024}. Next, we describe each step.

\subsection{Crash Report Grouping}\label{sec:crash_report_grouping}

Following related work~\cite{Wang2013, Wang2016, wu2018changelocator, medeiros2020, medeiros2024}, we group crash reports according to stack trace similarity to consolidate related failures. This grouping aims to reduce the effort required to identify the root causes of these failures. In the following discussion of grouping levels, we use the acronyms STA and STB to denote two stack traces in the crash report database.

{\bf (Level 1) Identical Stack Trace}. Initially, we group crash reports when two stack traces are identical (STA = STB). The signature representing each group is the stack trace itself.

{\bf (Level 2) Equivalent Signature}. After grouping identical stack traces, we further merge traces that differ only in minor implementation-specific details. For example, consider the following two stack traces, STA and STB:

\begin{itemize}
    \item \textit{at ...GeneratedMethodAccessor10184.invoke()} $\in$ STA 
    \item \textit{at ...GeneratedMethodAccessor10272.invoke()} $\in$ STB
\end{itemize}

\noindent These traces differ only in the generated numeric suffix -- \emph{10184} and \emph{10272}. Such variations do not affect the crash semantics; therefore, we treat the corresponding stack traces as equivalent and group their crash reports.

After handling exact and near-exact matches, we progressively broaden the similarity criteria to capture structurally related failures. The remaining levels follow the cumulative grouping strategy proposed by Wang et al.~\cite{Wang2016}.

{\bf (Level 3) Crash Type Signature}. We further merge Level 2 groups when one stack trace is structurally contained within another (STA $\subseteq$ STB or STB $\subseteq$ STA). To check containment, we compare only fully qualified method calls (package, class, and method names), ignoring line numbers, since they may vary due to formatting without affecting the crash semantics.

{\bf (Level 4) Top Frame File}. Finally, we merge Level 3 groups whose crash points originate from the same qualified top-frame file. For example, consider stack traces STA and STB, where:
\begin{itemize}
    \item \textit{at s.p.ClassMBean.methodA(ClassMBean.java:280)} is the signaler in STA 

    \item \textit{at s.p.ClassMBean.methodB(ClassMBean.java:251)} is the signaler in STB
\end{itemize}

In both cases, the method names differ, but the qualified file name is the same (\textit{s.p.ClassMBean}), and this rule groups such stack traces. In other words, if different stack traces signal exceptions in the same class, we consider them manifestations of the same underlying fault and group their crash reports, as validated in our previous work~\cite{medeiros2020, medeiros2024}.

\subsection{Suspicious file and method ranking}\label{sec:ranking_suspicious_files}

After grouping crash reports (Section~\ref{sec:crash_report_grouping}), we analyze the files appearing in each group's stack traces to identify those most likely responsible for the crash. To this end, we adapt the ranking approach proposed by Wu et al.~\cite{Wu} to the file level rather than the method level.
Each file is scored using three criteria:
(i) \gls{IAD} that favors class files that appear closer to the crash point in the stack trace;
(ii) \gls{IBF} that penalizes files frequently associated with different faults; and
(iii) \gls{FF} that rewards files that frequently appear within the same crash group.

The final score of a file $f$ appearing in the stack traces of a crash group $B$ is computed as follows:
\begin{equation}
Score(f, B) = \gls{IAD}(f,B)*\gls{IBF}(f, B)*\gls{FF}(f,B)\label{Score}
\end{equation}

The multiplicative formulation ensures that highly ranked files simultaneously satisfy three complementary properties: proximity to the crash point (IAD), specificity to the crash group (low IBF), and recurrence within the same failure context (FF). This combination prioritizes artifacts that are both structurally close to the failure signal and statistically consistent within the crash cluster, while penalizing overly generic files frequently associated with unrelated faults. Files are therefore ranked in descending order of this score to derive structured and prioritized contextual information, highlighting the most plausible root-cause candidates while reducing noise from unrelated crash groups.

{\bf \textit{Suggesting suspicious methods}}. After ranking files, we identify the methods within the top-ranked files that appear in the stack traces of the crash group and suggest them according to their frequency.

\subsection{Bug fixing}\label{sec:bug_fixing_stb}

After ranking suspicious files and methods for a crash report group, we extract the source code of the top-ranked files together with a representative stack trace to construct a structured contextual prompt. This prompt is submitted to the \gls{llm} to generate a patch suggestion, which is then attached to the corresponding crash report to assist developers during bug triage and bug resolution.

Our prompt design builds upon the findings of Du et al.~\cite{du2023resolving,du2026exploring}, who showed that enriched crash information (e.g., exception type and error message) and role-based instructions improve LLM performance in bug localization and repair. Following their insights, we adopt a zero-shot strategy and structure the prompt into five components (Listing~\ref{lst:zeroshot_prompt}): role specification, suspicious source code files, crash information (stack traces or stack messages), ranking information, and a repair request. We use zero-shot prompting due to input size constraints and context window limitations of current \glspl{llm}. Additionally, zero-shot prompting avoids potential overfitting to handcrafted exemplars and better reflects realistic industrial deployment scenarios, where representative labeled examples may not be readily available for each new crash category.

\lstset{breaklines=true}
\begin{lstlisting}[
  caption={Zero-shot prompt structure},
  label={lst:zeroshot_prompt},
  mathescape=true,
  %backgroundcolor = \color{lightgray!20},
  frame = l,
  basicstyle=\ttfamily\footnotesize,
  xleftmargin={0.65cm},
  breakindent=0.5em,
  numbers=left
]
$\textbf{Role-Play-Part}$: I want you to act as a fault localization and program repair expert. You will be able to provide the solutions' source code to fix the given program crash. Do not provide any explanations, just source code.
$\textbf{Suspicious-Code-Part}$: This is my code:
```
$\textit{[Suspicious file's source code 1..n]}$
```
$\textbf{Stack-Trace-Information-Part}$: I am getting:
```
$\textit{[Stack trace 1..n]}$
```
$\textbf{Stack-Message-Information-Part}$: I am getting:
```
$\textit{[Stack message 1..n]}$
```
$\textbf{Ranking-Information-Part}$: This is the ranking of classes and methods most suspected of containing the bug:
$\textit{1) Class1.java:}$
$\textit{  - methodA}$
$\textit{3) Class2.java:}$
$\textit{  - methodB}$
$\textit{  - methodC}$
$\textit{n) ClassN.java:}$
$\textit{  - methodN}$
$\textbf{Asking-Part}$: Give me the patched code to fix the program crash.
\end{lstlisting}

\section{Empirical Study} \label{sec:empirical_study_stb}

This study evaluates the integration of crash report clustering, suspicious file and method ranking, and \gls{llm}-based patch suggestion in an industrial setting. We investigate whether structured contextual information derived from crash mining can effectively support \gls{llm}-based bug localization and repair. To guide this investigation, we address the following research questions:

\emph{\rqastb} This question analyzes the form and content of LLM outputs, including code completeness, granularity, and explanatory behavior.

\emph{\rqbstb} This research question investigates whether and which information produced by the clustering approach contributes to improving the bug fixes generated by \glspl{llm}. We investigate this question through an ablation-style evaluation that systematically varies exception granularity and ranking information to quantify their marginal contributions to localization and repair effectiveness. This ablation study involves two stages: (i) initially on a subset of crash bugs using different LLMs and prompt configurations; and (ii) then evaluating whether the localization and repair performance observed in the initial experimental subset remains consistent when applied to the complete dataset.

Following the evaluation guidelines for empirical studies in Software Engineering involving \glspl{llm}~\cite{baltes2025evaluation}, Section~\ref{sec:stb_target_systems} describes the target systems and industrial context. Section~\ref{sec:target_llms_stb} details the selected LLMs and configurations. Section~\ref{sec:stb_prompt_combinations} presents the prompt combinations. Section~\ref{sec:stb_metrics} describes the evaluation procedure and human validation.

\subsection{Target Systems} \label{sec:stb_target_systems}
The study was conducted in collaboration with the software development department of a public institution responsible for maintaining several large-scale web-based systems. We focused on two Java Enterprise systems that together receive over one million daily requests and comprise approximately 2.1 million lines of code across more than 12,000 classes (Table~\ref{tab:stb_systems_overview}). Both systems have been in operation for over 16 years and are actively used and customized by more than 30 institutions. Approximately 70\% of the codebase consists of Java files, 29\% corresponds to web pages (HTML, JSP, JSF), and the remaining 1\% consists of other assets (build files, CI/CD configuration files, and so on).

\begin{table}[!t]\centering
    \caption{Target Systems Characterization}
    \label{tab:stb_systems_overview}
    \begin{tabular}{lrrrr}
        \hline
        System & Classes & Lines of code & Daily Requests \\ 
        \hline
        SIGAA & 8,156 & 1,323,196 & 1,208,326 \\
        SIGRH & 4,405 & 778,588 & 92,877  \\
        \hline
        Total & 12,561 & 2,101,784 & 1,301,203  \\
        \hline
    \end{tabular}
\end{table}

The first system (SIGAA) in Table~\ref{tab:stb_systems_overview} is an Integrated Academic Management System and generates approximately 9,200 crash reports per week. SIGRH is an Integrated Human Resource Management System, generating approximately 290 weekly crash reports.

The organization involved in this study does not maintain enterprise-level agreements with \gls{llm} providers that grant unrestricted source code access, mainly due to intellectual property and cost constraints. This may also be the case for many companies of similar size and characteristics. Therefore, the use of \glspl{llm} was limited to REST API calls with restricted input, providing only the source code of a few suspicious files to balance cost and confidentiality. This constraint underscores the need to provide LLMs with structured and prioritized contextual information.

\subsection{Target \glspl{llm}} \label{sec:target_llms_stb}

We selected \glspl{llm} based on the following criteria: 
(i) widespread adoption; 
(ii) availability via REST API; 
(iii) cloud-based execution support; and 
(iv) context window size.
Thus, we evaluated the following models:
\begin{itemize}
\item \texttt{claude-3-5-sonnet-20240620} (Anthropic), 
\item \texttt{gemini-1.5-pro} (Google), 
\item \texttt{gpt-4o-mini} (OpenAI), 
\item \texttt{mistral-large-2407} (Mistral), and 
\item \texttt{llama3-405b-instruct-maas} (Meta).
\end{itemize}

To avoid bias introduced by restrictive parameter tuning, we fixed the temperature at 1.0 and retained the default provider configurations unless mandatory parameters were required (e.g., \texttt{max-tokens=8192} for Anthropic).

\subsection{Prompt combinations} \label{sec:stb_prompt_combinations}

All prompt configurations are built upon the source code of the top-ranked suspicious files (SCoSF), enriched with crash-related contextual information. In particular, we systematically vary two components of the context provided to the \gls{llm}s: (i) the level of exception detail---either the full stack trace (ST) or only the stack message (SM); and (ii) the inclusion of structured ranking information, namely the Rank of Suspicious Files and Methods (RoSFM) produced by our clustering and ranking approach.

The first dimension controls the granularity of exception information supplied to the model, ranging from the complete execution context (ST) to a reduced representation containing only the exception type and error message (SM). Stack messages (SM) therefore consist only of exception types and their corresponding messages in a stack trace, including nested causes introduced by the \textit{``Caused by:''} lines, and exclude the full method call stack. This reduction significantly decreases prompt size and cost, allowing us to assess whether full stack traces are necessary for effective repair. The second dimension evaluates whether explicitly providing ranked suspicious artifacts (RoSFM) improves localization and repair performance compared to relying solely on raw crash information.

Based on these factors, we evaluated the following prompt configurations:
\begin{itemize}
\item \textbf{SCoSF+ST}: source code of suspicious files and the full stack trace.
\item \textbf{SCoSF+ST+RoSFM}: source code, stack trace, and ranking of suspicious files and methods.
\item \textbf{SCoSF+SM}: source code and stack message (exception type and error message only).
\item \textbf{SCoSF+SM+RoSFM}: stack message combined with ranking information.
\end{itemize}

In all configurations, we provide the source code of the top three suspicious files identified by our ranking approach. As mentioned earlier, providing the full codebase would be impractical in our setting due to confidentiality constraints, costs, rate limits, and the context window limitations of LLMs. Crash information is included to guide the LLM toward resolving the specific failure without addressing unrelated issues.

The resulting configurations enable an ablation analysis of contextual enrichment components, allowing us to quantify the marginal contribution of ranking information and exception granularity to \gls{llm}-based crash bug localization and repair.

\subsection{Evaluation metrics} \label{sec:stb_metrics}
We adopted a manual evaluation protocol, using the developer’s corrective commit as ground truth to assess whether the LLM-generated responses correctly localized the modified artifact and produced a semantically valid repair. For example, a repair was considered correct (plausible) if the LLM modified the same method as the developer and introduced a change that would prevent the observed exception, even if the implementation differed from the developer’s patch. This decision stems from limited test infrastructure, restricted access to full source artifacts, high rebuilding costs, and the need to manually assess semantic equivalence, as LLM-generated patches may differ syntactically from developer fixes while still resolving---or failing to resolve---the underlying fault. Given these constraints, automated metrics such as \textit{pass@k}~\cite{chen2021evaluating} or BLEU~\cite{papineni2002bleu} were unsuitable, since correctness depended on semantic validation rather than syntactic similarity or test-based execution.

Accordingly, we evaluated performance at two levels of granularity: (1) per-response, capturing correctness across all individual attempts; and (2) per-bug, capturing consistency across multiple attempts for the same crash bug. The \textit{Per-response} assessment involves two metrics: \textit{Localization@response} and  \textit{Repair@response}, where 

\begin{enumerate}[(a)]
\item {\textit{Localization@response}} measures the proportion of responses that correctly identify the artifact modified in the developer’s commit:

\begin{equation} \label{eq:Localization@response}
Localization@response = \frac{\sum_{i=1}^M localization(r_i)}{M} 
\end{equation}

\item \textit{Repair@response} measures the proportion of responses that produce a valid patch equivalent to, or a valid alternative to, the developer’s fix:

\begin{equation} \label{eq:Repair@response}
Repair@response = \frac{\sum_{i=1}^M repair(r_i)}{M} 
\end{equation}

\end{enumerate}
In both $@response$ metrics, $M$ denotes the total number of responses, and the indicator functions return 1 for correct localization or repair and 0 otherwise.

To assess robustness under repeated attempts, we evaluate correctness at the bug level by considering consistency across multiple responses. A bug is considered successfully localized (or repaired) if the majority of its responses are correct.

Formally, for a bug $b$ with $N$ attempts, we define a binary success indicator based on a majority threshold $T$. Accordingly, we leverage two indicators in the \emph{Per-bug} assessment: \textit{Localization@attempts} and \textit{Repair@attempts}, where:

\begin{equation}\label{eq:Localization@attempt_b}
\scalebox{0.88}{$
Localization@attempts(b) =
\begin{cases}
1 & \text{if } \sum_{j=1}^{N} localization(r_{b,j}) \geq T \\
0 & \text{otherwise}
\end{cases}
$}
\end{equation}

\begin{equation}\label{eq:Repair@attempt_b}
Repair@attempts(b) =
\begin{cases}
1 & \text{if } \sum_{j=1}^{N} repair(r_{b,j}) \geq T \\
0 & \text{otherwise}
\end{cases}
\end{equation}

The final per-bug metrics are obtained by averaging these indicators across all bugs $B$:

\begin{equation}\label{eq:Localization@bug}
Localization@bug = \frac{1}{|B|} \sum_{b \in B} Localization@attempts(b)
\end{equation}

\begin{equation}\label{eq:Repair@bug}
Repair@bug = \frac{1}{|B|} \sum_{b \in B} Repair@attempts(b)
\end{equation}

\subsection{Study Procedures}\label{stb_study_procedures}

\begin{figure*}[!t]
    \centering
    \includegraphics[width=0.9\textwidth]{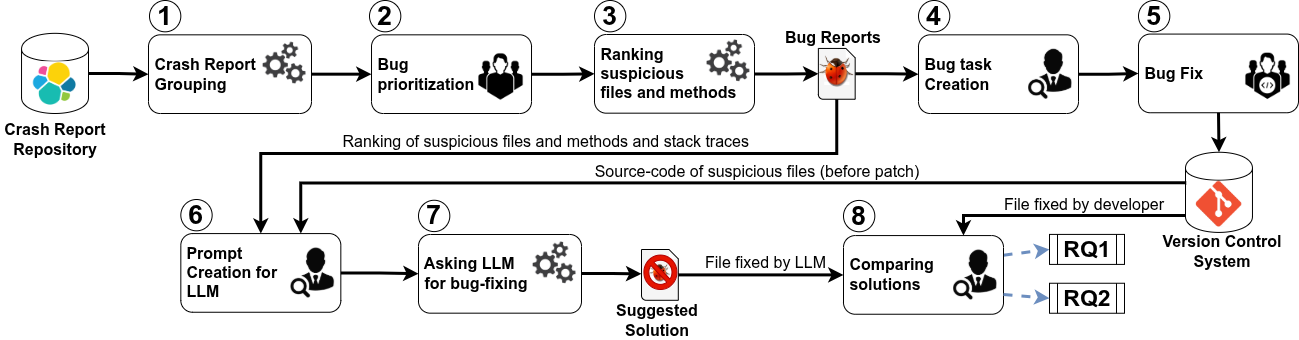}
    \caption{Study overview}
    \label{fig:stb_fig_flow}
\end{figure*}

We conducted (i) a longitudinal industrial deployment of the crash mining and bug management pipeline, and (ii) a retrospective evaluation of \gls{llm}-based localization and repair using artifacts generated during real maintenance activities. Fig.~\ref{fig:stb_fig_flow} provides an overview of the study workflow.

\paragraph{Industrial Phase}
Over an 18-month period, we executed the crash mining and bug management process on a weekly basis in the production environment. This phase comprises steps (1)--(5) in Fig.~\ref{fig:stb_fig_flow}.

In the first step \textbf{(Step 1 --- Crash report grouping)}, we collected crash reports generated during the previous seven days from the production systems and clustered them using the approach described in Section~\ref{sec:crash_report_grouping}. For each crash group, we shared the following information with the development teams: the group identifier, the first and last occurrence dates, the number of crash reports, the number of affected \glspl{URI}, the total number of impacted users, and the list of system classes appearing in the stack traces. Next, in the second step \textbf{(Step 2 --- Bug prioritization)}, the development teams analyzed this information to assess the impact and scope of each crash group and selected those to be addressed in the current development cycle.

After that, in the third step \textbf{(Step 3 --- Ranking suspicious files and methods)}, for each selected crash group, we extracted the files and methods appearing in the stack traces and ranked these assets according to the approach described in Section~\ref{sec:ranking_suspicious_files}.
In the fourth step \textbf{(Step 4 --- Bug task creation)}, we created issue-tracker tasks containing the top five suspicious files and methods together with additional contextual information extracted from the crash-report database.
In the final step \textbf{(Step 5 --- Bug fixing)}, developers were assigned these tasks, identified the root cause, implemented the corresponding patch, and closed the issue.

After executing these five steps continuously for 18 months, we constructed a dataset containing the buggy code, the ranking of suspicious files and methods, the most frequent stack traces, and the patched code for each crash group. More details are provided in Section~\ref{sec:dataset_overview_stb}.

\paragraph{Retrospective LLM Evaluation}
This phase comprises steps (6)--(8) in Fig.~\ref{fig:stb_fig_flow}. In the sixth step \textbf{(Step 6 --- Prompt creation)}, for each item in our dataset we instantiated the prompt template detailed in Section~\ref{sec:bug_fixing_stb}, including at most the top three \gls{SCoSF} and the most frequent \gls{ST} of the group. This restriction reflects practical \gls{llm} context-window and rate-limit constraints typically observed in industrial settings. To conduct an ablation-style analysis, we generated four prompt configurations, as detailed in Section~\ref{sec:stb_prompt_combinations}. For each crash, we generated all prompt combinations and used them in every attempt across all \glspl{llm}. Consequently, there was no difference in the content sent to the \glspl{llm}. This retrospective setup enables controlled comparison with developer fixes while approximating realistic debugging scenarios encountered in industrial maintenance.

Next, in the seventh step \textbf{(Step 7 --- Asking \gls{llm} for bug fixing)}, we submitted each prompt to the corresponding \gls{llm} via REST API using the configuration described in Section~\ref{sec:target_llms_stb}. To mitigate cross-request contamination, we executed each invocation in a fresh session with no shared conversational history. Given the stochastic nature of \gls{llm} outputs, we performed five individual zero-shot requests per prompt configuration. For an initial subset of the dataset, we evaluated five distinct \glspl{llm} (Section~\ref{sec:target_llms_stb}) for each prompt combination to answer \rqastbnumber\ and \rqbstbnumber. After that, we applied the best-performing \gls{llm}, using all prompt combinations, to the full dataset to assess robustness and answer the second stage of \rqbstbnumber.

Finally, in the eighth step \textbf{(Step 8 --- Comparing solutions)}, we compared each \gls{llm} response with the corresponding developer solution and analyzed the correctness of localization and repair. Two reviewers independently assessed whether the response correctly localized the modified artifact and whether it produced a semantically valid repair. Disagreements were resolved through discussion, and a third reviewer made an additional judgment in cases where the first two reviewers did not reach consensus.

Localization was considered correct when the \gls{llm} suggested a fix to the same snippet of code that the developer changed in the bug fix commit. For \texttt{NullPointerException}, we judged localization as correct only if the \gls{llm} fix attempted to resolve the problem in the same variable identified by the developer. Modifying the correct file, but a different code location counts as incorrect. The prompts averaged 2,244 lines of code (range: 100–7,717).

Repair was considered correct (plausible) when the generated patch was equivalent to, or a valid alternative implementation of, the developer's fix. Unsuccessful repair responses were categorized as: (i) ``Incomplete fix'', (ii) ``Provided a fix but introduced a new bug'', (iii) ``Tried to fix something else'', (iv) ``Incomplete response'', (v) ``Suggested a fix but not the patched code'', or (vi) ``Requested more details''. This classification scheme was inspired by Sobania et al.~\cite{sobania2023analysis}.

\section{Results and Discussion} \label{sec:results_and_discussion_stb}
This section presents the results of our dataset construction process and our analysis of the structural and explanatory characteristics of the \gls{llm} responses. It also reports the information provided by the clustering and ranking approach, which assists the \gls{llm} in locating and fixing the bugs that led to crash reports. Finally, it examines whether the effectiveness of the best-performing \gls{llm} remains consistent when evaluated across the full dataset.

\subsection{Dataset Building} \label{sec:dataset_overview_stb}

We built our dataset based on crash bugs grouped, ranked, and prioritized by our localization approach and subsequently fixed by development teams (Steps~1--5 in Fig.~\ref{fig:stb_fig_flow}). Over 18 months, 131 correction issues were opened on crash groups. Teams autonomously analyzed 86 (prioritized by demand and criticality): 24 closed without crash reproduction, 62 produced code changes, and 50 had Java commits. We selected crash groups that were resolved through code changes and for which a corrective commit could be clearly identified in the version history. From an initial set of 14 SIGAA and 31 SIGRH issues, we sampled ~20\% per system to form \sds. One SIGRH issue was excluded during its analysis. Later, while analyzing the remaining issues, 4 SIGAA and 2 SIGRH issues were excluded for the same reason (exceeded \gls{llm} context limits). We mined GitLab for the buggy (before the commit) and patched (after the commit) versions of the top three suspicious files for the resulting 38 resolved issues (10 from SIGAA and 28 from SIGRH), allowing a fair comparison between \gls{llm}-suggested fixes and developer patches. For each issue, we also included the top three ranked suspicious files and the top two frequent stack traces to complete the dataset, with the developer-modified file ranked within the top three for all 38 issues (36 first, 1 second, 1 third). We limited the number of files and stack traces due to \gls{llm} API limits. All issues were code-related, with no environmental factors involved.

\subsection{Analysis of LLM Responses and Effectiveness} \label{sec:analysis_llm_responses_stb}

To evaluate \rqastbnumber\ and the first stage of \rqbstbnumber, we conducted a full cross-model comparison on the eight-issue \sds\ (3 from SIGAA and 5 from SIGRH) described in Section~\ref{sec:dataset_overview_stb}. Each crash bug required the manual assessment of 100 \gls{llm} outputs (4 prompt combinations × 5 models × 5 attempts), resulting in 800 responses for the \sds alone. Expanding this analysis to all 38 crash bugs would require evaluating 3,800 responses, which would be infeasible in the context of this study. We therefore adopted \sds for model comparison and subsequently validated the best-performing configuration on the full dataset (\fds) to assess robustness.

Each LLM-generated patch was independently assessed by at least two authors. Disagreements were resolved by consensus, with no need for a third reviewer. Inter-rater agreement, measured via Cohen's Kappa~\cite{medcalc_kappa}, was substantial for bug localization ($\kappa$=0.767, 95\% CI [0.723, 0.812]) and moderate for both repair status ($\kappa$=0.500, 95\% CI [0.425, 0.576]) and fix category ($\kappa$=0.583, 95\% CI [0.544, 0.622]).

\noindent{\bfseries{\rqastb}} \label{sec:rqa_stb_results}
After analyzing the responses generated by \glspl{llm}, we observed that they varied considerably in structure. Responses included one or more entire classes, partial fragments, or a mixture of both, as illustrated in Listing~\ref{lst:class_fragments_response}. To preserve confidentiality, the listings were anonymized by modifying method and variable names.

Even when the prompt explicitly instructed the model to return only source code without any explanation, \glspl{llm} frequently violated this constraint---providing rationales either as inline comments within the code or as free text outside the code blocks, as shown in Listing~\ref{lst:class_fragments_response}. Furthermore, \glspl{llm} provided instructions for completing the code in various ways (\textit{\ldots}, \textit{//\ldots}, \textit{// \ldots (existing code)}, \textit{//\ldots (rest of the method)}, among others). We also observed the omission of documentation in the suggested code. These omissions, combined with non-compliance with prompt instructions, the diversity of responses, and the inherent non-determinism of \glspl{llm}, represent a relevant challenge for automated patch extraction pipelines in settings similar to ours.

\lstset{breaklines=true}
\begin{lstlisting}[
  caption={Fragments Single Class},
  label={lst:class_fragments_response},
  frame = l,
  basicstyle=\ttfamily\footnotesize,
  xleftmargin={0.65cm},
  breakindent=0.5em,
  numbers=left
]
The bug is in the `isDateVisible` method of the `AbsenceMBean` class. This method is called in the `form.jsp` file, and it throws a `NullPointerException` because the `attachment` object is null.
To fix this bug, we need to ensure that the `attachment` object is not null before calling the `isDateVisible` method. We can do this by adding a null check in the `isDateVisible` method:
```java
public boolean isDateVisible(){
    if (attachment == null) {
        return false;
    }
    return !isEmpty(attachment.getDocType()) && attachment.getDocType().isDateRequired();
}
```
Alternatively, we can also initialize the `attachment` object in the `startUpdate` method before calling the `isDateVisible` method:
```java
public String startUpdate() throws ArqException {
    // ...
    attachment = new absenceDocument();
    attachment.setDocType(new DocType());
    // ... (rest of the method)
}
```
\end{lstlisting}

Therefore, we manually classified the responses to characterize them and analyze the feasibility of using automated methods to measure effectiveness. We classified the responses into the following types: \textit{Full Single Class}, \textit{Full Multiple Classes}, \textit{Full Class + Partial Class}, \textit{Fragment Multiple Classes}, \textit{Fragment Single Class}, \textit{Fragments Single Class}, \textit{Non-Java Code}, and \textit{Explanation Only}.

Table~\ref{tab:llm_structural_response_characteristics} shows that a single fragment of one class was the most frequent response type (\textit{Fragment Single Class}), considering every \gls{llm} and prompt. In the tables, "all" denotes the aggregate across LLMs, prompts, or both.
It also reveals differences across prompt combinations (rows~2--5).
\textit{Fragment Single Class} was most frequent in prompts containing stack traces, while prompts formed only by \gls{SCoSF} and \gls{SM} generated more responses with the full source code of multiple classes (\textit{Full Multiple Classes}). Adding \gls{RoSFM} to prompts with \gls{SM} led \glspl{llm} to more frequently suggest multiple fragments of a single class (\textit{Fragments Single Class}). Analyzing responses by \gls{llm} (rows~6--10), \texttt{Claude} generated more \textit{Fragments Single Class}, \texttt{Gemini} and \texttt{Mistral} responded more with \textit{Full Multiple Classes}, while \texttt{GPT} and \texttt{Llama} more often produced \textit{Fragment Single Class}. In particular, only \texttt{Mistral} generated \textit{Non-Java Code} and \textit{Explanation Only} responses. \texttt{Gemini} and \texttt{Mistral} also returned the full source code of a single class (\textit{Full Single Class}), although these responses were usually incomplete or truncated.

\begin{table*}[!t]\centering
    \caption{Structural response characteristics (\sds)}
    \label{tab:llm_structural_response_characteristics}
    \setlength{\tabcolsep}{3pt}
        \begin{tabular}{llrrrrrrrr}
            \hline
            & & Full & Full & Full class + & Fragment & Fragment & Fragments & Non-Java & Explanation \\
            \gls{llm} & Prompt & Single Class & Multiple Classes & Partial class & Multiple Classes & Single Class & Single Class & Code & Only \\
            \hline

            all & all & 2.63\% & 26.75\% & 0.13\% & 4.00\% & {\bfseries 34.00\%} & 26.38\% & 4.00\% & 2.13\% \\
            \hline

            \multirow{4}{*}{all} & \gls{SCoSF}+\gls{ST}        & 2.00\% & 24.50\% & 0.00\% & 1.50\% & {\bfseries 44.00\%} & 21.50\% & 6.00\% & 0.50\% \\
            & \gls{SCoSF}+\gls{ST}+RoSFM                        & 3.50\% & 28.00\% & 0.50\% & 2.50\% & {\bfseries 36.50\%} & 22.50\% & 3.00\% & 3.50\% \\
            & \gls{SCoSF}+\gls{SM}                               & 1.00\% & {\bfseries 33.00\%} & 0.00\% & 5.00\% & 29.50\% & 27.50\% & 2.50\% & 1.50\% \\
            & \gls{SCoSF}+\gls{SM}+RoSFM                        & 4.00\% & 21.50\% & 0.00\% & 7.00\% & 26.00\% & {\bfseries 34.00\%} & 4.50\% & 3.00\% \\
            \hline

            claude  & \multirow{5}{*}{all} & 0.00\% & 6.25\%  & 0.00\% & 7.50\% & 36.88\%             & {\bfseries 49.38\%} & 0.00\%  & 0.00\%  \\
            gemini  &                      & 0.63\% & {\bfseries 80.63\%} & 0.00\% & 0.00\% & 0.00\%  & 0.00\%              & 18.75\% & 0.00\%  \\
            gpt     &                      & 0.00\% & 1.25\%  & 0.00\% & 3.75\% & {\bfseries 70.63\%} & 24.38\%             & 0.00\%  & 0.00\%  \\
            mistral &                      & 11.88\% & {\bfseries 44.38\%} & 0.63\% & 3.13\% & 10.00\% & 18.13\%            & 1.25\%  & 10.63\% \\
            llama   &                      & 0.63\% & 1.25\%  & 0.00\% & 5.63\% & {\bfseries 52.50\%} & 40.00\%             & 0.00\%  & 0.00\%  \\
            \hline
        \end{tabular}
\end{table*}

Documentation suppression in the source occurred in 43\% of responses, including Javadoc comments and inline annotations. Additionally, explanations outside code blocks appeared in 45.5\% of responses, while 37.73\% contained explanations embedded within the generated code. Furthermore, 15.88\% of responses suppressed documentation and added explanations both within and outside the source code blocks.

\texttt{Claude} exhibited the most pronounced tendency---suppressing documentation in 68.13\% of its responses, adding external explanations in 90.63\%, and adding internal explanations in 51.88\%. In contrast, \texttt{Gemini} was the only model that neither suppressed documentation nor added unsolicited explanations. However, it produced the highest rate of truncated responses (48.75\%).

These results highlight the difficulty of automatically generating an adequate patch to the buggy code. Due to this difficulty and others discussed in Section~\ref{sec:stb_metrics}, we proceeded by manually analyzing the \gls{llm} suggestions to answer \rqbstbnumber. 

\begin{rqanswer}{Answer to \rqastbnumber}
\noident
\gls{llm} responses vary substantially across models and prompts and often omit documentation or add unsolicited explanations, complicating automated patch extraction.
\end{rqanswer}

\noindent{\bfseries{\rqbstb}} \label{sec:rqb_stb_results}
To answer this research question, we first assessed how many \gls{llm} responses referenced at least one method changed by the developer to fix the bug. \textit{{\bfseries Analyzing the \sds}}. Overall, 
65.75\% of responses included such a method---either in a code block or in an explanation. Prompts with \gls{ST} generated the most references to developer-patched methods (76\%), and adding \gls{RoSFM} further increased this rate to 81\%, suggesting that ranking information helps \glspl{llm} focus on the relevant code location. In contrast, combinations with \gls{SM} yielded lower rates (38\%--68\%). Analyzing results by model rather than by prompt, \texttt{Claude} performed best (78\%), with \texttt{Llama} (75\%) and \texttt{GPT} (72\%) exceeding the average, while \texttt{Gemini} (53\%) and \texttt{Mistral} (51\%) performed below average.

These results suggest that \glspl{llm} frequently identify the same methods that developers modified to fix the crash. Building on this observation, we further investigated whether \glspl{llm} could localize and repair bugs at a finer granularity---specifically, within individual methods.

For the \textit{Per-response granularity}, we first evaluate each approach’s ability to identify bugs in individual responses.
Table~\ref{tab:llm_localization_repair_accuracy} (Columns~2--5) reports localization accuracy per response (Section~\ref{sec:stb_metrics}). \texttt{Claude-3.5-sonnet} achieved the highest overall localization accuracy (82.5\%) with \gls{SCoSF}+\gls{ST}+\gls{RoSFM}, and nearly 70\% with \gls{SCoSF}+\gls{SM}+\gls{RoSFM}. \texttt{Gemini} followed the same pattern, peaking with \gls{SCoSF}+\gls{ST}+\gls{RoSFM}, while \texttt{GPT} tied between \gls{SCoSF}+\gls{ST} and \gls{SCoSF}+\gls{ST}+\gls{RoSFM}. \texttt{Mistral} achieved its best result with \gls{SCoSF}+\gls{SM}+\gls{RoSFM} and \texttt{Llama3} with \gls{SCoSF}+\gls{ST}. Considering repair accuracy (Columns~6--9), \texttt{Claude} again achieved the best overall performance, although its best prompt shifted to \gls{SCoSF}+\gls{SM}+\gls{RoSFM} (52.5\%), which is 7.5 percentage points higher than \gls{SCoSF}+\gls{ST}+\gls{RoSFM} (45.0\%).
\texttt{Gemini} mirrored this pattern, while \texttt{GPT} tied between \gls{SCoSF}+\gls{ST} and \gls{SCoSF}+\gls{SM}. \texttt{Mistral} peaked with \gls{SCoSF}+\gls{ST}+\gls{RoSFM}, and \texttt{Llama3} with \gls{SCoSF}+\gls{ST}.

\begin{table*}[!t]
\centering
\setlength{\tabcolsep}{4pt}
\caption{Per-Response Localization and Repair Accuracy (\sds)}
\label{tab:llm_localization_repair_accuracy}
\begin{tabular}{l|rrrr|rrrr}
\hline
 & \multicolumn{4}{c|}{Localization@response} & \multicolumn{4}{c}{Repair@response} \\
\cline{2-5} \cline{6-9}
\gls{llm}
& \gls{SCoSF}+\gls{ST}
& \gls{SCoSF}+\gls{ST}
& \gls{SCoSF}+\gls{SM}
& \gls{SCoSF}+\gls{SM}
& \gls{SCoSF}+\gls{ST}
& \gls{SCoSF}+\gls{ST}
& \gls{SCoSF}+\gls{SM}
& \gls{SCoSF}+\gls{SM}
\\

& 
& +\gls{RoSFM}
& 
& +\gls{RoSFM}
& 
& +\gls{RoSFM}
& 
& +\gls{RoSFM} 
\\

\hline
claude
& 30/40 (75.0\%)
& {\bfseries 33/40 (82.5\%)}
& 15/40 (37.5\%)
& 27/40 (67.5\%)
& 20/40 (50.0\%)
& 18/40 (45.0\%)
& 9/40 (22.5\%)
& {\bfseries 21/40 (52.5\%)} \\

gemini
& 12/40 (30.0\%)
& 15/40 (37.5\%)
& 10/40 (25.0\%)
& 10/40 (25.0\%)
& 3/40 (7.5\%)
& 0/40 (0.0\%)
& 1/40 (2.5\%)
& 4/40 (10.0\%) \\

gpt
& 14/40 (35.0\%)
& 14/40 (35.0\%)
& 10/40 (25.0\%)
& 13/40 (32.5\%)
& 10/40 (25.0\%)
& 9/40 (22.5\%)
& 10/40 (25.0\%)
& 9/40 (22.5\%) \\

mistral
& 14/40 (35.0\%)
& 11/40 (27.5\%)
& 10/40 (25.0\%)
& 15/40 (37.5\%)
& 8/40 (20.0\%)
& 10/40 (25.0\%)
& 8/40 (20.0\%)
& 9/40 (22.5\%) \\

llama3
& 16/40 (40.0\%)
& 14/40 (35.0\%)
& 10/40 (25.0\%)
& 10/40 (25.0\%)
& 10/40 (25.0\%)
& 6/40 (15.0\%)
& 5/40 (12.5\%)
& 6/40 (15.0\%) \\

\hline
\end{tabular}
\end{table*}


For the \textit{per-bug granularity} assessment, we measured the number of solved problems, following Sobania et al.~\cite{sobania2023analysis}. However, we considered a bug solved when at least three of five responses produced a correct patch. Unlike their automated test-based verification, we performed manual validation due to the lack of test suites and the difficulty of rebuilding the systems.
Table~\ref{tab:llm_crash_group_localization_repair} reports localization and repair rates at the bug level (at least 3 of 5 responses correct).

\texttt{Claude} localized bugs consistently in 87.5\% of issues with \gls{SCoSF}+\gls{ST}+\gls{RoSFM}, and achieved 50\% repair across three prompt combinations. \texttt{GPT} and \texttt{Mistral} tied across prompts for repair (25\%), while \texttt{Llama3} peaked with \gls{SCoSF}+\gls{ST} (25\%). These variations across models and metrics reinforce that prompt selection is model-dependent and should be considered when fine-tuning \gls{llm}-based repair pipelines.

\begin{table*}[!t]
\centering
\setlength{\tabcolsep}{9pt}
\caption{Per-Bug Localization and Repair Accuracy (\sds)}
\label{tab:llm_crash_group_localization_repair}
\begin{tabular}{l|rrrr|rrrr}
\hline
 & \multicolumn{4}{c|}{Localization@bug} & \multicolumn{4}{c}{Repair@bug} \\
\cline{2-5} \cline{6-9}
\gls{llm}
& \gls{SCoSF}+\gls{ST}
& \gls{SCoSF}+\gls{ST}
& \gls{SCoSF}+\gls{SM}
& \gls{SCoSF}+\gls{SM}
& \gls{SCoSF}+\gls{ST}
& \gls{SCoSF}+\gls{ST}
& \gls{SCoSF}+\gls{SM}
& \gls{SCoSF}+\gls{SM}
\\

& 
& +\gls{RoSFM}
& 
& +\gls{RoSFM}
& 
& +\gls{RoSFM}
& 
& +\gls{RoSFM} 
\\
\hline

claude
& 6/8 (75.0\%)
& {\bfseries 7/8 (87.5\%)}
& 3/8 (37.5\%)
& 5/8 (62.5\%)
& {\bfseries 4/8 (50.0\%)}
& {\bfseries 4/8 (50.0\%)}
& 2/8 (25.0\%)
& {\bfseries 4/8 (50.0\%)} \\

gemini
& 2/8 (25.0\%)
& 3/8 (37.5\%)
& 2/8 (25.0\%)
& 2/8 (25.0\%)
& 0/8 (0.0\%)
& 0/8 (0.0\%)
& 0/8 (0.0\%)
& 1/8 (12.5\%) \\

gpt
& 3/8 (37.5\%)
& 3/8 (37.5\%)
& 2/8 (25.0\%)
& 2/8 (25.0\%)
& 2/8 (25.0\%)
& 2/8 (25.0\%)
& 2/8 (25.0\%)
& 2/8 (25.0\%) \\

mistral
& 3/8 (37.5\%)
& 2/8 (25.0\%)
& 2/8 (25.0\%)
& 3/8 (37.5\%)
& 2/8 (25.0\%)
& 2/8 (25.0\%)
& 2/8 (25.0\%)
& 2/8 (25.0\%) \\

llama3
& 3/8 (37.5\%)
& 3/8 (37.5\%)
& 2/8 (25.0\%)
& 2/8 (25.0\%)
& 2/8 (25.0\%)
& 1/8 (12.5\%)
& 1/8 (12.5\%)
& 1/8 (12.5\%) \\

\hline
\end{tabular}
\end{table*}


Table~\ref{tab:classes_of_llm_answers} presents the response classification for the best-performing \gls{llm}. Including \gls{RoSFM} in the prompt appears to increase the number of fixes similar to developer patches. Overall, \textit{\gls{SCoSF}+\gls{SM}+\gls{RoSFM}} (without \gls{ST}) achieved the highest repair rate (52.5\%), while prompts using \gls{SM} instead of \gls{ST} more often attempted to fix unrelated issues.

\begin{table*}[!t]
    \centering
    \caption{Categories of \gls{llm}-Generated Fixes for claude-3.5-sonnet-20240620 (\sds)}
    \label{tab:classes_of_llm_answers}
    \begin{tabular}{lrrrr}
        \hline
        Category & \gls{SCoSF}+\gls{ST} & \gls{SCoSF}+\gls{ST}+\gls{RoSFM} & 
        \gls{SCoSF}+\gls{SM} & \gls{SCoSF}+\gls{SM}+\gls{RoSFM} \\ 
        \hline
        Fix similar to the developer's & 25.0\% & 32.5\% & 15.0\% & 37.5\% \\
        Fix using an alternative implementation & 25.0\% & 12.5\% & 7.5\% & 15.0\% \\
        Incomplete fix & 7.5\% & 7.5\% & 0.0\% & 0.0\% \\
        Provides a fix but introduced a new bug & 17.5\% & 30.0\% & 15.0\% & 15.0\% \\
        Tries to fix something else & 25.0\% & 17.5\% & 62.5\% & 32.5\% \\
        Incomplete response & 0.0\% & 0.0\% & 0.0\% & 0.0\% \\
        Suggested a fix but not the patched code & 0.0\% & 0.0\% & 0.0\% & 0.0\% \\
        \hline
    \end{tabular}
\end{table*}

Despite the three-way tie in Repair@bug (Table~\ref{tab:llm_crash_group_localization_repair}), \gls{SCoSF}+\gls{SM}+\gls{RoSFM} stands out as the best overall prompt combination because it achieved the highest repair rate (52.5\%) and the lowest rate of unsuccessful fixes---encompassing incomplete fixes, patches that introduced new bugs, and attempts targeting unrelated issues (Table~\ref{tab:classes_of_llm_answers})---making it the most reliable configuration in the \sds. 

\noindent\rule{\linewidth}{0.4pt}\\
\noindent\textbf{Insight.} \textit{The \gls{llm} that most strictly followed the prompt instructions (\texttt{Gemini}) consistently underperformed, while the one that most frequently added unsolicited explanations (\texttt{Claude}) achieved the best results.}
\rule{\linewidth}{0.4pt}

\begin{rqanswer}{Answer to \rqbstbnumber \,\, (on the \sds)}
\noident
Most \glspl{llm} correctly referenced or attempted to fix at least one developer-patched method, particularly when prompts included stack traces and the ranking of suspicious files and methods. \texttt{Claude-3.5-sonnet} achieved the best localization performance (82.5\% accuracy; 87.5\% at-least-3-of-5) with \gls{SCoSF}+\gls{ST}+\gls{RoSFM}, and the best repair results (52.5\% accuracy; 50.0\% at-least-3-of-5) with \gls{SCoSF}+\gls{SM}+\gls{RoSFM}.
\end{rqanswer}

\begin{table*}[!ht]\centering
    \caption{Localization and Repair Accuracy (\fds) --- Claude-3.5-sonnet-20240620}
    \label{tab:full_dataset_llm_accuracy}
    \setlength{\tabcolsep}{3pt}
    \begin{tabular}{lrr|rr}
        \hline
        & \multicolumn{2}{c|}{Per-response} & \multicolumn{2}{c}{Per-bug} \\
        \cline{2-3} \cline{4-5}
        Prompt 
        & Localization@response
        & Repair@response
        & Localization@bug
        & Repair@bug \\
        \hline
        all
        & 473/760 (62.24\%) & 332/760 (43.68\%)
        & 29/38 (76.32\%)   & 22/38 (57.89\%) \\
        \hline
        \gls{SCoSF}+\gls{ST}
        & \textbf{135/190 (71.05\%)} & \textbf{98/190 (51.58\%)}
        & \textbf{27/38 (71.05\%)}   & \textbf{20/38 (52.63\%)} \\
        \gls{SCoSF}+\gls{ST}+\gls{RoSFM}
        & 132/190 (69.47\%) & 90/190 (47.37\%)
        & 26/38 (68.42\%)   & 17/38 (44.74\%) \\
        \gls{SCoSF}+\gls{SM}
        & 84/190 (44.21\%)  & 54/190 (28.42\%)
        & 17/38 (44.74\%)   & 11/38 (28.95\%) \\
        \gls{SCoSF}+\gls{SM}+\gls{RoSFM}
        & 122/190 (64.21\%) & 90/190 (47.37\%)
        & 24/38 (63.16\%)   & 18/38 (47.37\%) \\
        \hline
    \end{tabular}
\end{table*}

\textit{{\bfseries Analyzing the \fds.}} After identifying the best-performing \gls{llm} on the \sds (21\% of issues), we evaluated \texttt{Claude-3.5-sonnet} on the \fds (38 crash bugs). 
Inter-rater agreement was almost perfect for bug localization ($\kappa$=0.882, 95\% CI [0.832, 0.908]) and moderate for both repair status ($\kappa$=0.507, 95\% CI [0.446, 0.568]) and fix category ($\kappa$=0.454, 95\% CI [0.416, 0.493]).
Disagreements were resolved by consensus; a third reviewer was consulted as a tiebreaker for one response.
In general, the proportion of responses that referenced at least one developer-patched method decreased slightly from 78\% to 76\%, with \textit{\gls{SCoSF}+\gls{SM}} being the only combination that improved (45\% to 49\%).

\setcounter{paragraph}{0}
With respect to \textit{Per-response granularity}, 
Table~\ref{tab:full_dataset_llm_accuracy} reports per-response accuracy on the \fds. For localization, \textit{\gls{SCoSF}+\gls{ST}} achieved the best result (71.05\%), while \textit{\gls{SCoSF}+\gls{ST}+\gls{RoSFM}}---the top combination in the \sds---dropped from 82.5\% to 69.47\%. For repair, \textit{\gls{SCoSF}+\gls{ST}} also achieved superior performance (51.58\%), with the largest gain observed for \textit{\gls{SCoSF}+\gls{SM}} (22.5\% to 28.42\%), while the previously best-performing combination \textit{\gls{SCoSF}+\gls{SM}+\gls{RoSFM}} declined from 52.5\% to 47.37\%.

Considering the assessment at the bug level (i.e., \textit{Per-bug granularity} assessment), Table~\ref{tab:full_dataset_llm_accuracy} shows that the combinations with \gls{ST} generally declined for localization, while those with \gls{SM} improved. For localization, \textit{\gls{SCoSF}+\gls{ST}} remained the top performer (71.05\%), despite dropping from 75\%, while \textit{\gls{SCoSF}+\gls{ST}+\gls{RoSFM}} decreased from 87.5\% to 68.42\%. For repair, \textit{\gls{SCoSF}+\gls{ST}} was the only combination to improve (50\% to 52.63\%), while \textit{\gls{SCoSF}+\gls{SM}+\gls{RoSFM}} dropped from 50\% to 47.37\% and \textit{\gls{SCoSF}+\gls{SM}} gained modestly (25\% to 28.95\%).

Analysis of fix categories (Table~\ref{tab:full_dataset_classes_of_llm_answers}) corroborates these findings: \textit{\gls{SCoSF}+\gls{ST}} achieved the highest correct fix rate (51.58\%) and the lowest incomplete fixes, bug introductions, and unrelated fix (48.42\%) among all combinations. Notably, this contrasts with the \sds results, where \textit{\gls{SCoSF}+\gls{SM}+\gls{RoSFM}} was the top combination.

\begin{table*}[!t]\centering
    \caption{Categories of \gls{llm}-Generated Fixes for claude-3.5-sonnet-20240620 (\fds)}
    \label{tab:full_dataset_classes_of_llm_answers}
        \begin{tabular}{lrrrr}
            \hline
            Category & \gls{SCoSF}+\gls{ST} & \gls{SCoSF}+\gls{ST}+\gls{RoSFM} & 
            \gls{SCoSF}+\gls{SM} & \gls{SCoSF}+\gls{SM}+\gls{RoSFM} \\ 
            \hline
            Fix similar to the developer's & 43.68\% & 41.58\% & 24.21\% & 40.53\% \\
            Fix using an alternative implementation & 7.89\% & 5.79\% & 4.21\% & 6.84\% \\
            Incomplete fix & 4.21\% & 3.68\% & 2.63\% & 2.11\% \\
            Provides a fix but introduced a new bug & 15.26\% & 18.42\% & 13.16\% & 14.74\% \\
            Tries to fix something else & 28.95\% & 28.42\% & 55.26\% & 35.79\% \\
            Incomplete response & 0.00\% & 0.00\% & 0.00\% & 0.00\% \\
            Suggested a fix but not the patched code & 0.00\% & 2.11\% & 0.00\% & 0.00\% \\
            Requested more details & 0.00\% & 0.00\% & 0.53\% & 0.00\% \\
            \hline
        \end{tabular}
\end{table*}

Although adding \textit{\gls{RoSFM}} slightly reduced performance when \textit{\gls{ST}} was provided, it substantially improved \textit{\gls{SCoSF}+\gls{SM}} performance: per-response localization (44.21\% to 64.21\%) and repair (28.42\% to 47.37\%), and per-bug localization (44.74\% to 63.16\%) and repair (28.95\% to 47.37\%).

\begin{rqanswer}{Answer to \rqbstbnumber \,\, (on the \fds)}
\noident
Using \texttt{Claude-3.5-sonnet} on the full dataset, the approach generalized well, with \textit{\gls{SCoSF}+\gls{ST}} emerging as the best combination for both localization (71.05\%) and repair (52.63\%), with \gls{RoSFM}'s benefit being context-dependent (helping with \gls{SM}, not with \gls{ST}).
\end{rqanswer}

\section{Threats to Validity} \label{sec:threats_to_validity_stb}
Following established guidelines for empirical software engineering studies~\cite{wohlin2012experimentation, runeson2009guidelines}, we discuss threats to internal, construct, conclusion, and external validity.

\textit{{\bfseries Internal Validity.}}
Buggy and patched code were extracted from Redmine issues and Git commits, which may contain incomplete or imprecise information. Although fixes are validated by a quality assurance team before issue closure, in the industrial context studied some commits may include partial fixes or unrelated changes. To mitigate this risk, we manually inspected commits and associated issue discussions when identifying the ground-truth artifact. Another threat concerns the manual evaluation of \gls{llm}-generated responses. Each response was independently analyzed by two authors, with disagreements resolved by a third reviewer.

\textit{{\bfseries Construct Validity.}}
Effectiveness was measured using the metrics \textit{Localization@response}, \textit{Repair@response}, \textit{Localization@bug}, and \textit{Repair@bug}, which assess whether \glspl{llm} correctly locate the faulty artifact and produce a valid repair relative to the developer’s commit. This operationalization relies on manual inspection of \gls{llm}-generated responses rather than automated execution or test-based validation, which may introduce subjectivity when assessing semantic equivalence between generated patches and developer fixes. To reduce this threat, responses were independently evaluated by two reviewers, with arbitration by a third reviewer when necessary. Additionally, data leakage is unlikely to affect our measurements: both systems reside on a private Git server accessible only via the company's internal network, and neither the source code nor the developer fixes are publicly available, making leakage of the bugs and fixes evaluated highly unlikely.

\textit{{\bfseries Conclusion Validity.}}
Because \glspl{llm} are non-deterministic, responses may vary across executions. To mitigate this threat, we issued five independent requests per prompt configuration and considered a bug successfully localized or repaired only when a majority of responses were correct. Nevertheless, randomness in generation may still influence the results.

\textit{{\bfseries External Validity.}}
Our study analyzes crash reports from two Java-based enterprise systems developed by the same company, which may limit generalization to other technologies or domains. Furthermore, the evaluation was conducted retrospectively using historical crash reports and developer fixes, and the results may differ in real-time debugging scenarios where developers interact with \glspl{llm}. Finally, we evaluated only general-purpose \glspl{llm}; different results may arise with specialized models or future versions of these systems.

\section{Related Work} \label{sec:related_work_stb}

Du et al.~\cite{du2023resolving, du2026exploring} investigated the use of \glspl{llm} for resolving crash bugs and showed that interactive strategies, such as role-play prompts and multi-round interactions, improve effectiveness, particularly for code-related crashes. Their approach (IntDiagSolver) uses small Stack Overflow snippets, achieving high localization but modest repair results. In contrast, our study focuses on industrial crash reports, uses zero-shot prompts with richer context (full suspicious files, stack traces or messages, and ranked suspicious artifacts), and evaluates larger, real-world codebases. While our localization accuracy is slightly lower (71\% vs.\ 76.7\%), our repair accuracy is higher (52\% vs.\ 33.3\%).

Sobania et al.~\cite{sobania2023analysis} evaluated \texttt{ChatGPT} for automatic bug fixing on the QuixBugs dataset, showing competitive performance with state-of-the-art approaches and highlighting the benefits of dialog-based interaction. They considered a bug solved if at least one of four responses succeeded, correctly repairing 19 of 40 bugs (47.5\%). They also classified responses into multiple categories, noting that many of them required additional information or failed to identify the bug. Our study adopts a stricter success criterion (at least 3 correct fixes out of 5 responses) and a single interaction per response. Using richer contextual prompts on real-world industrial Java crash bugs, we achieved 71\% localization and 52\% repair accuracy with \texttt{Claude-3.5-sonnet} --- comparable repair performance despite the harder setting.

Fahim et al.~\cite{fahim2025crash} enhanced crash reports from issue trackers of open-source systems by enriching them with stack trace fragments, method-level source code, and \gls{llm}-generated fixes, achieving up to 58\% localization and 41\% repair accuracy using an agentic strategy. In contrast, our work operates on raw crash reports automatically collected in production from industrial systems, without textual descriptions. We cluster crashes and rank suspicious files and methods based on stack traces, then provide this structured context to \glspl{llm} to suggest fixes. Evaluating multiple models and prompt combinations, we found that \texttt{Claude-3.5-sonnet} achieved the best performance, localizing 71\% of crash bugs and correctly repairing 52\% of them in an industrial setting --- competitive results considering the rawer, less structured nature of our crash data.

\section{Conclusion} \label{sec:conclusion_stb}

We investigated whether the use of \glspl{llm} can improve and complement a bug localization approach by providing development teams with suggestions to repair crash bugs in the source code. We evaluated the effectiveness of our bug-fixing approach retrospectively by analyzing real-world bugs in an industrial context involving large-scale web-based systems implemented with Java Enterprise technologies.

Initially, we analyzed 800 \gls{llm}-generated responses for 8 crash bugs to assess their structural and explanatory characteristics. We found that a single fragment from a single class is the most frequent response type and that both prompt combinations and the \gls{llm} model influence these characteristics.

We also compare the localization and repair performance of four prompt combinations across five \glspl{llm} on a subset of our dataset to identify the best-performing \gls{llm} and to analyze which information provided by our clustering and ranking approach helps \glspl{llm} locate and fix crash bugs. The \texttt{Claude-3.5-sonnet} model achieved better performance on bug localization tasks using prompts generated from the source code of suspicious files, stack trace samples, and the ranking of suspicious files and methods. The same model performed better on bug-fixing tasks when prompts used stack messages rather than stack traces.

Finally, we verify whether the effectiveness of the best-performing \gls{llm} generalizes to the full dataset (38 crash bugs). We achieved better performance in bug localization (71\%) and repair (52\%) tasks using prompts composed exclusively of the source code of suspicious files and stack trace samples. These results indicate that structured crash-mining context can enable LLMs to effectively support bug localization and repair in industrial maintenance workflows.

\section*{Acknowledgment}

Many thanks to INES.IA (www.ines.org.br), CNPq (408817/2024-0, 311749/2025-94-0), CAPES (88887.186324/2025-00), and STI/UFRN for partial support.

\bibliographystyle{IEEEtran}
\bibliography{IEEEabrv,references}

\end{document}